\documentclass[trackchanges]{aastex701}
\usepackage{graphicx}%
\usepackage{subcaption}
\usepackage{multirow}%
\usepackage{amsmath,amssymb,amsfonts}%
\usepackage{amsthm}%
\usepackage{booktabs}%

\begin{document}

\title{On the Blueprint of Active Galaxies Producing Neutrinos}

\correspondingauthor{Chen Li}

\author[orcid=0009-0000-4848-5113,sname='Li']{Chen Li}
\affiliation{University of Wisconsin–Madison, Department of Physics}
\email[show]{chenli@icecube.wisc.edu}  

\author[orcid=0000-0001-6224-2417,sname='Halzen']{Francis Halzen}
\affiliation{University of Wisconsin–Madison, Department of Physics}
\email[show]{halzen@icecube.wisc.edu}

\begin{abstract}

Based on the observation of the active galaxies NGC 1068 and TXS 0506+056, and on additional evidence for the  sources NGC 4151, CGCG 420-015, NGC 7469, and the Circinus Galaxy emerging from IceCube data, we make the case for the production of high-energy neutrinos within a few gravitational radii of supermassive black holes surrounded by a dense plasma radiating X-rays. X-rays represents the target for the production of neutrinos by protons accelerated near the black hole; they also absorb the gamma rays from the decay of neutral pions produced in the same interactions. Neutrinos with energies of tens of TeV and above originate in photoproduction interactions with X-rays of $0.1 \sim 1$\,keV energy on the $\Delta$ resonance, $p + \gamma \rightarrow \Delta \rightarrow n + \pi^+$. Our analysis of the multimessenger data points to gamma-ray-obscured sources with a characteristic neutrino flux linearly proportional to the X-ray flux originating within $\sim 10$ gravitational radii of the black holes, with lower values preferred. We speculate on such sources producing the diffuse flux of neutrinos and cosmic rays of extragalactic origin.

\end{abstract}




\section{Introduction}\label{sec1}

The active galaxy NGC 1068 has been identified as a source of high-energy neutrinos with a pre(post)-trial-corrected significance of $5.3(4.2)\sigma$\,\citep{IceCube:2022der}. With 100 neutrinos, it is also the hottest spot in the neutrino sky and aligns with its astronomical position to $0.04^\circ$ after 13 years of data\,\citep{Abbasi:2025tas}. As a galaxy with abundant star formation, it had been anticipated to be a high-energy neutrino source\,\citep{Bykov:2020zqf}, but this interpretation of the IceCube data can be ruled out by the fact that the gamma rays from $\pi^0$ decay that accompany the charged pions producing neutrinos are not observed. The observation underscored the fact that a powerful neutrino emitter requires a dense target to efficiently transfer the energy of accelerated protons into a neutrino flux; this target is likely to absorb these pionic gamma rays, as is indeed the case for NGC 1068. By the same argument, jets transparent to high-energy gamma rays are unlikely to be efficient neutrino emitters.

The dense core of plasma surrounding the core of its central black hole, dubbed the corona, has turned out to provide a compelling alternative site for neutrino production. Protons accelerated near the black hole interact with the abundant X-ray flux it radiates in $p\gamma$ interactions; at the same time, it efficiently absorbs the pionic gamma rays to evade the strong upper limits established by the MAGIC ground-based gamma-ray telescope in the relevant energy range\,\citep{2022arXiv220203381M}.

Besides the corona, some active galaxies, including NGC 1068, host high densities of hydrogen clouds measured through their line-of-sight density, $\rm N_H$. Original modeling by the Penn State group\,\citep{murase2020hidden} predominantly produced neutrinos via $pp$ interactions, but still required the presence of the corona to efficiently absorb the accompanying gamma rays. In this paper, we investigate whether the simplified picture wherein the corona also represents the primary target to produce the neutrinos is viable. The possibility has been raised in the literature\,\citep{Halzen:2023usr,khatee2025ngc,Ding:2025tbd,Testagrossa:2026jcs}; here we will make a case for the dominant role of the X-rays in the corona as the neutrino-producing target by showing that this blueprint for producing neutrinos is supported by observations of TXS 0506+056 and by the evidence garnered from subleading sources emerging in the IceCube sky map. These observations are indeed consistently characterized by a high X-ray flux combined with the absence of high-energy gamma rays in the energy range in which they could have provided evidence for neutrino production.

We will introduce the corona model\,\citep{murase2020hidden} in more detail in Section \ref{sec2}. In Section \ref{sec3}, we will provide the rational for selecting, besides NGC 1068, the sources TXS 0506+056, NGC 4151, CGCG 420-015, the Circinus Galaxy, and NGC 7469 for further investigation in our proposal. We will argue that the multimessenger data of TXS 0506+056 can be accommodated in the same framework\,\citep{khatee2025ngc}. In Sections \ref{sec4} and \ref{sec5}, we will evaluate our proposal with the multimessenger data of all sources in more detail and show that an outstanding feature of the fits to the data is the preference for neutrino production near the central supermassive black hole, within $\sim 10$ gravitational radii or less. In the final discussion in Section \ref{sec5}, we will conclude by presenting evidence of a linear relation between the X-ray and neutrino flux of the source, previously argued for by Emma Kun and collaborators\,\citep{kun2024correlation}. We will also speculate on the possibility that this class of sources reproduce the diffuse flux of extragalactic neutrinos and cosmic rays.

\section{Introducing the Corona Model}\label{sec2}

We first remind the reader that IceCube's measurement of the diffuse high-energy neutrino flux~\citep{PhysRevLett.111.021103, icecube2013evidence, 
PhysRevLett.113.101101} provides key constraints on their source population(s). The IceCube Collaboration has refined its measurements of this signal, using larger datasets and improved event selection and reconstruction~\citep{aartsen2019measurements, aartsen2020characteristics, abbasi2021icecube, abbasi2022improved, abbasi2026time}. These studies have produced a spectrum that is broadly consistent with a power law with spectral index $\gamma \sim 2.3 - 2.9$. More recent analyses, however, indicate that the diffuse spectrum softens significantly in the TeV-PeV energy range and exhibits a feature at around $\sim 30$ TeV~\citep{abbasi2026evidence, abbasi2026improved}.

The spectral feature observed in the diffuse neutrino spectrum naturally arises from proton interactions with radiation proceeding through the $\Delta$-resonance\,\citep{KhateeZathul:2026oam}. Pions produced via the $\Delta$ resonance carry on average $\sim 20\%$ of the energy of the initial proton. Combined with the fact that each neutrino carries, on average, about one quarter of the pion energy in the decay chain $\pi^+ \rightarrow \mu^+ + \nu_\mu$ followed by $\mu^+ \rightarrow e^+ + \nu_e +\bar\nu_\mu$, this process yields neutrinos with $E_\nu \sim 0.05 \, E_p$. For interactions on the $\Delta$ resonance, the energy in the center-of-momentum frame satisfies the kinematic condition:
\begin{align}
m_{\Delta}^2 = E^2_{\rm CM} = 2 E_p E_{\gamma} +m^2_p - 2 E_{\gamma} \sqrt{E^2_p-m^2_p} \, \cos \theta,
\end{align}
where $\theta$ is the angle between the momenta of the proton and the target photon. Averaging over this angle, we obtain
\begin{align}
&m_{\Delta}^2 -m^2_p \approx  2 E_p E_{\gamma} \langle 1 - \cos \theta \rangle,
\end{align}
and thus
\begin{align}
E_p E_{\gamma} \approx \frac{m_{\Delta}^2 -m^2_p}{2} \approx 0.3 \, {\rm GeV}^2.
\end{align}
Producing a spectral feature at $E_{\nu} \sim 30 \, {\rm TeV}$, therefore, requires protons with energies $E_p \sim 0.6 \, {\rm PeV}$ and target photons in the $0.1-1 \, {\rm keV}$ range. Such a target is associated with the corona emission from dense cores in some active galaxies, an environment where multiple opportunities also exist for accelerating protons to produce neutrinos in interactions with the X-rays; see, e.g., Ref.~\citet{LeBihan:2026rqq}.

\begin{figure}[ht]
    \centering
    \includegraphics[width=0.75\columnwidth]{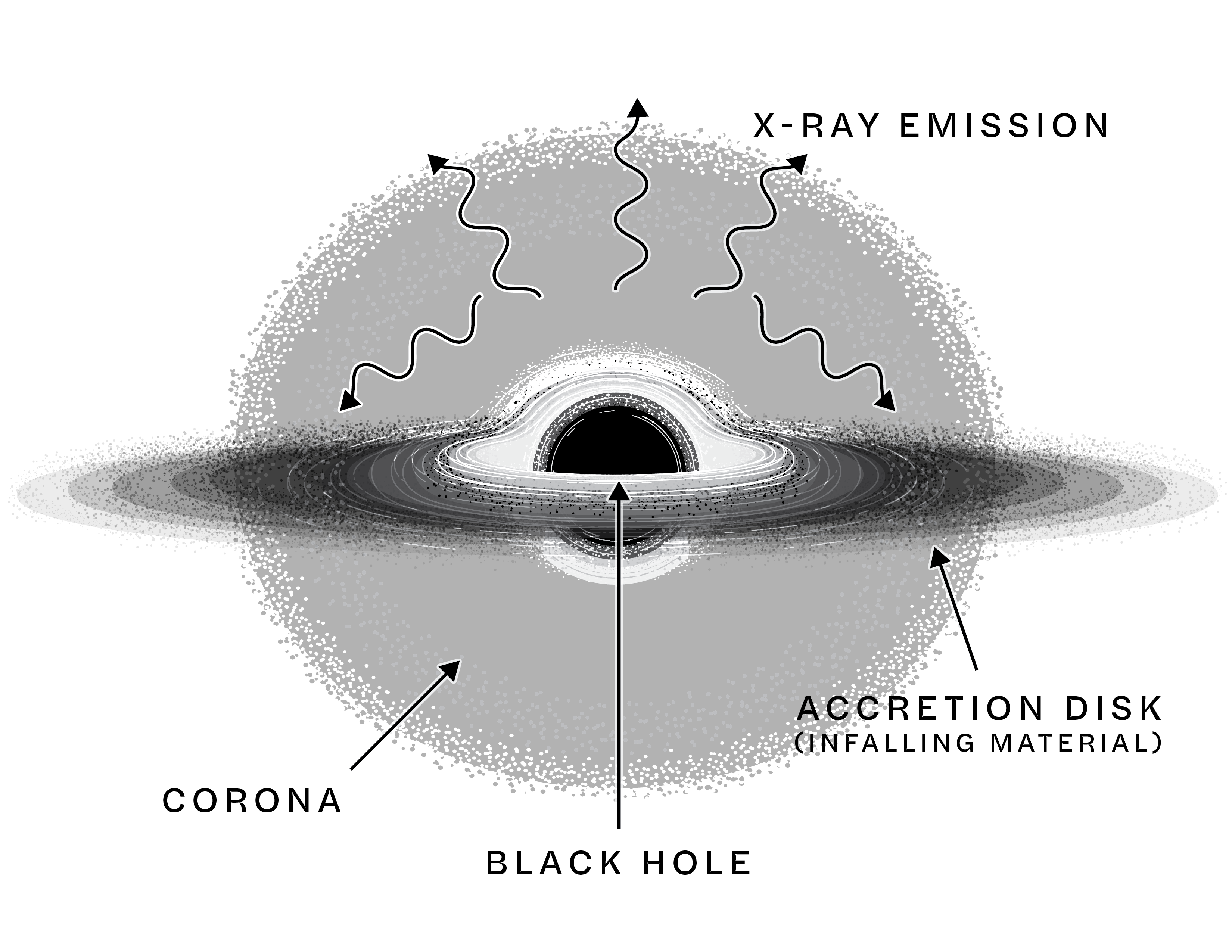}
    \caption{The core of an active galaxy combines the ingredients to produce neutrinos, with protons accelerated by the black hole and/or the accretion disk and a target (corona) with a large optical depth in both gamma rays and hydrogen that converts them into neutrinos.}
    \label{fig:Black-Hole_Diagram}
\end{figure}

The evidence is accumulating that neutrinos originate in $\gamma$-ray obscured sources; this should not come as a surprise. Imagine that protons, accelerated near the black hole or on the accretion disk, interact in a photon target of size $\rm R$ centered on the black hole; see Fig.~\ref{fig:Black-Hole_Diagram}. The opacity\footnote{The opacity is actually the energy-loss length, and $\tau$ should be replaced by $1-e^{-\tau}$ whenever it is large~\citep{Halzen:2022pez}.} for accelerated protons to interact in the target is
\begin{equation}
\label{eq:pgammaopacity}
\rm \tau_{p\gamma}\simeq \frac{\kappa_{p\gamma} R_{escape}}{\lambda_{p\gamma}} \simeq \kappa_{p\gamma} \rm R_{escape}\, \sigma_{p\gamma} \,n_{\gamma}.
\end{equation}
It is determined by the number of interactions of the proton in a target of size $\rm R_{target}$ given its interaction length $\lambda_{p\gamma}$. With each interaction, the proton loses a fraction $\kappa_{p\gamma}$ of its energy, the inelasticity which is $\sim 0.2$ in the $\Delta$-resonance approximation. The interaction length is determined by the density of target photons $n_\gamma$ and the interaction cross section $\sigma_{p\gamma}$\footnote{For the simple dimensional analysis in this section, we use the following cross sections: $\rm \sigma_{\gamma\gamma} = 6.7 \times 10^{-25} cm^2$, $\rm \sigma_{p\gamma} = 5 \times 10^{-28} cm^2$, and $\rm \sigma_{pp} = 3 \times 10^{-26} cm^2$.}.

The opacity of the target to the photons produced along with the neutrinos is given by
\begin{equation}
\tau_{\gamma\gamma} \simeq \rm R_{target}\, \sigma_{\gamma\gamma}\,n_{\gamma},
\end{equation}
and therefore, approximately, the two opacities are related by their cross sections
\begin{equation}
\tau_{\gamma\gamma} \simeq \frac{\sigma_{\gamma\gamma}}{\kappa_{p\gamma} \sigma_{p\gamma}} \, \tau_{p\gamma} \simeq 10^3\,  \tau_{p\gamma}
\label{opacitygamma}
\end{equation}
for $\rm R_{escape} \sim R_{target} \sim R$\footnote{There is an additional factor associated with the different thresholds of the two interactions; see Ref.~\citet{svensson1987non}.}. A target that produces neutrinos with $\tau_{p\gamma} \gtrsim 0.1$ will not be transparent to the pionic gamma rays, which will lose energy in the target even before propagating in the EBL. For instance, we should not expect neutrinos to be significantly produced in blazar jets that are transparent to very high-energy gamma rays. In contrast, the highly obscured dense cores close to supermassive black holes in active galaxies represent an excellent target for producing neutrinos in an environment with multiple opportunities for accelerating protons associated with the black hole or the turbulent magnetic fields above the accretion disk.

Pursuing these dimensional arguments, we estimate the corona opacity to protons from Eq.~\ref{eq:pgammaopacity} with a target density $n_X$ as
\begin{equation}
n_X = \frac{u_X}{E_X} = \frac{1}{4 \pi {\rm R}^2c}\, \frac{L_X}{E_X}\,,
\end{equation}
and therefore
\begin{equation}
\tau_{p\gamma} = \frac{\kappa_{p\gamma} \sigma_{p\gamma}}{4 \pi c}\, \frac{1}{\rm R}\, \frac {L_X}{E_X}.
\end{equation}
We here determined the target density in X-rays from the energy density of X-rays $\rm u_X$ divided by their energy $\rm E_X$ and, subsequently, identified the energy flux $\rm cu_X$ with the measured luminosity of the source $\rm 4 \pi R^2\, L_X$.

The optical depth can be rewritten in terms of the Eddington luminosity, $L_{\rm edd}$, and the Schwarzschild radius, $\rm R_S$, of a black hole of mass $M$:
\begin{eqnarray}
    \tau_{p\gamma}
    &\simeq&
    70 \,
    \Big(
    \frac{\rm R}{\rm R_S}
    \Big)^{-1}
    \Big(
    \frac{ E_{X} }{ 1 \text{ keV} }
    \Big)^{-1}
    \Big(
    \frac{ L_{X} }{ L_{\rm edd} }
    \Big)
    \,,
    \label{eq:tau}
\end{eqnarray}
This simplified picture of the core of an active galaxy reveals the main features of the corona model; with $L_\nu = \tau_{p \gamma}\,L_p$, where $L_p$ is the luminosity of accelerated protons in the source, we anticipate that the neutrino flux is proportional to the X-ray flux and inversely proportional to the radius, the two quantities that define the density of the target. This will emerge from the more detailed modeling of the formation of the the corona and an improved description of the photoproduction of pions in $p \gamma$ interactions.

We will show next that the corona model describes the multiwavelength data of NGC 1068 and of additional sources supported by IceCube data (without making a claim to discovery).

\section{Emerging IceCube High-Energy Neutrino Sources}\label{sec3}

In a routine IceCube data analysis, besides identifying the hottest spot(s) in the sky map, a search is performed for a predetermined list of 110 gamma-ray sources. Also, a binomial analysis is performed that matches the sky map incrementally, from two through 110 sources selected from the same list, and determines the most probable match. All analyses are performed blind using maximum likelihood techniques.

With a decade of data, such an analysis\,\citep{IceCube:2019cia} revealed four sources that emerged with a local p-value exceeding $3\sigma$: NGC 1068, PKS 1424+240, TXS 0506+056, and GB6 J1542+6129. The binomial fit selected three of the sources: NGC 1068, PKS 1424+240, and TXS 0506+056. A time-dependent search of the same data identified leading flares from PKS 1502+106, TXS 0506+056, M87, and GB6 J1542+6129. A later search with eight years of data exclusively obtained with the completed detector revealed a slightly different set of active galaxies near the $5\sigma$ discovery potential of the analysis: TXS 0506+056, 3C454.3, 4C 38.41, and Cygnus A\,\citep{IceCube:2018ndw}.

With improved detector calibrations and analysis tools applied to all, including archival, data, referred to as PASS 2 data, the global significance of NGC 1068 increased from $2.9$ to $4.2\sigma$\ post trial\,\citep{IceCube:2022der}. Its observation drew attention to the potential role of the X-ray luminosity in selecting neutrino sources, as discussed in the previous section. A dedicated search for X-ray bright sources found evidence for two additional sources: CGCG 420-015 and NGC 4151, supported by both a stacked and binomial analysis of the source list. A similar analysis\,\citep{IceCube:2026hzq} targeting the Southern Hemisphere found evidence for the Circinus Galaxy as a neutrino emitter. For this source, the KM3Net detector should eventually have improved sensitivity over what for IceCube represents a source in the Southern Hemisphere for which its sensitivity is reduced by a factor of $\sim 5$. Although KM3Net is only partially completed, it has already identified CGCG 420-015 as its leading source with a significance of $2.5\sigma$\,\citep{Celli}.

Finally, besides the list of 110 sources selected for their high-energy gamma-ray flux, the IceCube Collaboration preselected a list of 48 nonjetted sources with a bright X-ray core. Excluding NGC 1068, the binomial best fit to the new source list returned 11 sources with a local p-value of $3.3\sigma$. Some of the individual sources reached a local p-value of more than $2.5\sigma$: NGC 4151, CGCG 420-015, Cygnus A, and NGC 7469 reaching $3.8\sigma$. Although this is the first analysis where NGC 7469 makes an appearance, it had already been pinpointed as a neutrino source by the unique issuance of two IceCube alerts in its direction\,\citep{Sommani:2025yur}.

Among these sources TXS 0506+056 stands out because it had already been detected in 2017 in a multimessenger campaign involving some 20 multiwavelength observations. On September 22, 2017, IceCube reported a well-reconstructed muon that deposited 180 TeV inside the detector, corresponding to a most probable energy of 290\,TeV~\citep{kopper2017icecube,icecube2018neutrino} for the parent neutrino, known as IC-170922A. Its arrival direction was aligned with the coordinates of a known {\it Fermi} blazar, TXS~0506+056, to within $0.06^\circ$. The source was ``flaring" with a gamma-ray flux that had increased by a factor of seven in recent months. A variety of estimates converged on a probability on the order of $10^{-3}$ that the coincidence was accidental~\citep{icecube2018neutrino}. The identification of the neutrino with the source reached the level of evidence, but not more.  What clinched the association was a series of subsequent observations.

Searching archival IceCube data revealed that the neutrino luminosity of TXS~0506+056 is actually dominated by an earlier gamma-ray-obscured burst observed in 2014-2015~\citep{icecube2018neutrino} with a global p-value of $3.7\sigma$, which leaves the burst associated with IC170922 as a subdominant contribution. There is no evidence for radiation from TXS 0506+056, from optical to gamma rays, over the apparently hours-long event that produced IC170922.

The MASTER robotic optical telescope network, which had been monitoring the source since 2005, found its strongest time variation in the last 15 years to have occurred two hours after the emission of IC-170922, with a second less statistically compelling variation following the 2014 burst~\citep{lipunov2020optical}. The source switches from the ``off" to the ``on" state two hours after the emission of the neutrino. After monitoring the uniformity of their observations until the first quarter of 2020, they concluded that the time variation detected on September 22, 2017, is at a level of $50 \sigma$, which conclusively associates the source with the neutrino. Such strong variability may be caused by changes in the accretion rate of the central engine \citep{2023NatAs...7.1282R}. 

It is striking that all sources appear to be active galaxies with a dense X-ray core. Some are obscured in gamma rays in the energy range that corresponds to the decay products of neutral pions accompanying the neutrinos. Others, including TXS 0506+056 and GB6 J1542+6129, host high-energy gamma-ray jets but do not show evidence of gamma-ray emission during the emission of neutrinos\,\citep{khatee2025ngc}.

In this context, the selection of Cygnus A in the binomial analysis of the IceCube list of nonjetted X-ray sources, if confirmed, is interesting because its jets are not pointing in our direction, leaving its bright X-ray core as the obvious source of neutrinos. In contrast, the same analysis finds no evidence for neutrino emission from Centaurus A, despite the strong X-ray emission from jets that point in our direction.

In a more in-depth analysis of the multimessenger gamma-ray and radio observations of the active galaxy GB6 J1542+6129\,\citep{Kun:2026jbx}, evidence is presented for the spatial separation of the sites where neutrinos and gamma rays are produced. The observations are consistent with a disturbance-driven scenario in which neutrinos are produced in a compact, photon-rich region near the central engine, plausibly the corona or jet-base interface, while the same disturbance later enhances Doppler-boosted leptonic emission at the parsec-scale VLBI core. The delayed $gamma$-ray emission likely arises from Doppler-boosted leptonic processes in the jet. Recently, a similar separated but causally related feature has been discovered in the jet of TXS 0506-056 associated with IC170922\,\citep{Kovalev:2026fba}. Our emphasis that jetted galaxies may emit neutrinos from a dense X-ray core aligns with the concept of masquerading blazars introduced in Ref.~\citet{Padovani:2019xcv}.

In the end, we will select for further study only sources that have emerged from at least two independent analyses, leaving NGC 4151, CGCG 420-015, NGC 7469, and the Circinus Galaxy, besides NGC 1068 and TXS 0506+056. Despite the fact that more than 90 neutrinos have been associated with the strong source PKS 1425+240, it is not part of our list since an analysis has not been completed resolving the source confusion with NGC 5610\,\citep{Kun:2025cqc}. NGC 5610 is an X-ray bright Seyfert galaxy and thus a credible candidate neutrino emitter.

\section{Multimessenger Description of IceCube Sources: Method}\label{sec4}

We will next integrate the ideas presented above in more detailed simulations of the sources able to describe their observed neutrino, $\gamma$-ray, and X-ray fluxes. Our code is publicly available at\,\url{https://github.com/chenlinear/a2m3}.

\subsection{Simulation Setup}\label{sec41}

We first describe the setup of the simulations and introduce the parameters that are used to describe the multimessenger fluxes of the sources.

We adopt the corona model described in Section \ref{sec2}. Although the geometry of the corona is likely complicated, we simply model a sphere of volume $V = \frac{4 \pi}{3} R^3$, where $R$ is the radius, which is kept constant in the simulation. We assume that there is a homogeneous and isotropic magnetic field of strength $B$ and a uniform density of protons $n_p$ in the corona region. We further assume injection spectra uniform in time and space; they are composed of three parts:

\begin{itemize}
    \item Protons described by a power-law spectrum with an exponential high-energy cutoff:
    \begin{equation}
    \frac{dN_p}{dE_p dtdV} \begin{cases} \propto E_p^{-\alpha_p} \exp{(-E_p / E_{p, \mathrm{max}})}, & E_p \ge E_{p, \mathrm{min}} \\ = 0, & E_p < E_{p, \mathrm{min}} \end{cases},
    \end{equation}
    where $\alpha_p$ is the proton spectral index and $E_{p, \mathrm{min}}$ and $E_{p, \mathrm{max}}$ are the minimum and maximum proton energies, respectively. For convenience in sampling the parameter space, we will refer to them by their energy ratio $E_{p, \mathrm{max}} / E_{p, \mathrm{min}}$ and the minimum energy $E_{p, \mathrm{min}}$. The spectrum is normalized by the proton injection luminosity $L_p$ and the volume $V$. We do not implement an explicit acceleration mechanism. Other authors have done so for NGC 1068; see, for instance, Ref.~\citet{LeBihan:2026rqq}.
    \item A coronal electron plasma with an injection rate following a Maxwell-J\"uttner distribution\,\citep{tsouros2017energy}, which is a relativistic version of the Maxwell-Boltzmann distribution: 
    \begin{equation}
    \frac{dN_e}{dE_e dtdV} \propto \frac{\gamma_e^2 \beta_e}{\Theta_e K_2(1 / \Theta_e)}\exp{(-\gamma_e / \Theta_e)},
    \end{equation}
    where $E_e = \gamma_e m_e c^2$, $\beta_e = \sqrt{1 - 1 / \gamma_e^2}$, and $\Theta_e = T_e / m_e c^2$. Here, $T_e$ is the temperature of the electrons and $K_2$ the modified Bessel function of the second kind. As was done for protons, the spectrum is normalized to the electron injection luminosity $L_e$ and the volume $V$.
    \item Planck radiation\,\citep{dermer2009high} from the accretion disk: 
    \begin{equation}
    \frac{dN}{dE_{\gamma} dt dV} \propto \frac{8 \pi}{\lambda_\mathrm{C}^3} \frac{E_\gamma^2}{\exp{(E_\gamma / \Theta_\gamma)} - 1},
    \end{equation}
    where the electron Compton wavelength $\lambda_\mathrm{C} = h/m_e c = 2.42 \times 10^{-10} \, \mathrm{cm}$ and $\Theta_\gamma = T_\gamma / m_e c^2$. Here, $T_\gamma$ is the temperature of the photons. Similarly, this spectrum is normalized to the photon injection luminosity $L_\gamma$ and volume $V$. Note that the flux of Planck radiation is proportional to volume; therefore, $L_\gamma$ defines the effective volume of the disk for the corona. The X-rays produced by electron plasma and Planck radiation through the inverse Compton (IC) process and other processes is what constitutes the corona.
\end{itemize}

We inject all particle spectra into the simulation code Astrophysical Multimessenger Modeling (AM3)\,\citep{klinger2024am3} with all processes activated, namely, synchrotron radiation, IC, photon-photon pair annihilation, Bethe-Heitler pair production, photo-pion process, inelastic proton-proton collisions, and pion and muon decay. We set a global escape timescale of all particle species to be $t_{\mathrm{esc}} = {\rm R}/c$.

After simulation using the AM3 software, the energy variation with redshift and the absorption of photons on the extragalactic background light (EBL)\,\citep{dominguez2011extragalactic} are taken into account\footnote{We assume a flat $\Lambda$-CDM universe with Hubble constant $H_0 = 70 \, \mathrm{km/s/Mpc}$ and matter density $\Omega_{m, 0} =  0.3$.}.

\subsection{Dataset Generation}\label{sec42}

To sample the parameter space, we generate a Latin hypercube parameter set, the configurations of which are outlined in Table~\ref{tab:parameter_configs}. These parameters are used for both runs A and B, with the additional parameter $n_p$ introduced in run B, which varies according to the values specified. The benefits of using Latin hypercube sampling are that it can better cover the parameter space and that there is no need to manually set the bin size of each parameter. With AM3, we simulate all the points in this parameter space to obtain a neutrino and photon flux. After initializing the kernel, we empty all the particle fluxes, put in the injection spectra, and simulate. When we move on to the next point in this parameter space, we empty all the particle fluxes but do not initialize the kernel again unless the previous parameter point crashes the kernel, in which case we skip this parameter point.

\begin{table}[htbp]
\centering
\caption{Parameter configurations used to generate the Latin hypercube parameter set. Note that $T_e$ and $T_\gamma$ are fixed to the values indicated.}
\label{tab:parameter_configs}
\begin{tabular}{lcll}
\toprule
\multicolumn{1}{c}{\textbf{Parameter}} & \textbf{Distribution} & \textbf{Low} & \textbf{High} \\
\midrule
$B$ \hfill [G] & Logarithmic & $10^{3}$ & $10^{9}$ \\
$R$ \hfill [cm] & Logarithmic & $10^{12}$ & $10^{15}$ \\
$L_p$ \hfill [erg/s] & Logarithmic & $10^{42}$ & $10^{51}$ \\
$E_{p, \mathrm{min}}$ \hfill [eV] & Logarithmic & $10^{12}$ & $10^{15}$ \\
$E_{p, \mathrm{max}} / E_{p, \mathrm{min}}$ & Logarithmic & $1$ & $10^{3}$ \\
$\alpha_p$ & Linear & $1.5$ & $6$ \\
$L_e$ \hfill [erg/s] & Logarithmic & $10^{37}$ & $10^{44}$ \\
$T_e$ \hfill [keV] & - & \multicolumn{2}{c}{$500$} \\
$L_\gamma$ \hfill [erg/s] & Logarithmic & $10^{37}$ & $10^{44}$ \\
$T_\gamma$ \hfill [eV] & - & \multicolumn{2}{c}{$50$} \\
$n_p$ \hfill [cm$^{-3}$] & Logarithmic & $10^{-1}$ & $10^{12}$ \\
\bottomrule
\end{tabular}
\end{table}

We did two runs: run A and run B. They share all assumptions and ranges of parameters,  except that run A exclusively assumes $p \gamma$ interactions with $n_p = 0$, while run B allows for $p \gamma$ and $pp$ interactions with $n_p$, the density of protons, as a free parameter. To compensate for the added parameter $n_p$, we increase the run B simulations by a factor of $8$, balancing efficiency when generating the dataset. Table~\ref{tab:runs} summarizes a comparison of runs A and B.

\begin{table}[htbp]
\centering
\caption{Information contrasting modeling runs A and B.}
\label{tab:runs}
\begin{tabular}{lcc}
\toprule
 & \textbf{Run A} & \textbf{Run B} \\
\midrule
Total Number of Simulations & $10,485,760$ & $83,886,080$ \\
Number of Free Parameters & 8 & 9 \\
CPU Hours & $\sim 7,000$ & $\sim 80,000$ \\
Success Rate & $98.69 \%$ & $98.17 \%$ \\
\bottomrule
\end{tabular}
\end{table}

\begin{table}[htbp]
\centering
\caption{Properties of the active galaxies modeled.}
\label{tab:source}
\begin{tabular}{lcc}
\toprule
 & $\rm R_S$ [cm] & $d_L$ [Mpc] \\
\midrule
NGC 1068 & $3.9 \times 10^{12}$ \citep{padovani2024high} & $14.4$ \\
TXS 0506+056 & $9.3 \times 10^{12}$ \citep{tjus2022neutrino} & $1774$ \citep{paiano2018redshift} \\
NGC 4151 & $5 \times 10^{12}$ \citep{bentz2022broad} & $15.8$ \citep{yuan2020cepheid} \\
CGCG 420-015 & $5.9 \times 10^{13}$ \citep{tanimoto2018suzaku} & $128.8$ \citep{ricci2017bat} \\
Circinus Galaxy & $5 \times 10^{11}$ \citep{greenhill2003warped} & $4.1$ \\
NGC 7469 & $2.66 \times 10^{12}$ \citep{bentz2015agn} & $70$ \\
\bottomrule
\end{tabular}
\end{table}

\subsection{Observations}\label{sec43}

We are now ready to fit the neutrino, $\gamma$-ray, and X-ray fluxes of the selected sources: NGC 1068, TXS 0506+056, NGC 4151, CGCG 420-015, the Circinus Galaxy, and NGC 7469. Table~\ref{tab:source} describes the properties of the active galaxies modeled. The goal of the simulation is to describe the neutrino flux observed, while for the X-ray and $\gamma$-ray fluxes we consider the data as upper limits because their flux originates from multiple processes. Where the neutrino flavors are concerned, following our assumptions, the neutrino and antineutrino flavor ratios are $\nu_e : \nu_\mu : \nu_\tau = 1 : 2 : 0$ at the source. Because the cooling of pions produced by $p \gamma$ and $pp$ is small as they decay quickly, we obtain a composition of $\nu_e : \nu_\mu : \nu_\tau = 1 : 1 : 1$ at observation given the long oscillation length.

For NGC 1068, IceCube data, 4FGL-DR2 data, and MAGIC data are from \citet{IceCube:2022der}. Other multiwavelength observations are from the Markarian Multiwavelength Data Center (MMDC)\,\citep{sahakyan2024markarian}. We use 4FGL-DR2, MAGIC, BAT105m \citep{oh2018105}, and RASS \citep{boller2016second} data for filtering the simulated photon fluxes in Section \ref{44}. For TXS 0506+056, the neutrino data is from \citet{IceCube:2022der}, the photon observations are from the MMDC dataset, and we use 2FHL \citep{ackermann2016fermi}, MMDCGR, MMDCXRT, and NuBlazar \citep{middei2022first} data when filtering the simulated photon fluxes. For NGC 4151, the neutrino data is from \citet{abbasi2025search} and the $\gamma$-ray data from \citet{murase2024sub}. For CGCG 420-015, the neutrino data is from \citet{abbasi2025search} and the $\gamma$-ray data from \citet{ma2025seyfert}. For the Circinus Galaxy, neutrino data is from \citet{abbasi2026evidence2} and the $\gamma$-ray data from \citet{murase2024sub}. For NGC 7469, the neutrino data is from \citet{abbasi2026evidence2} and the $\gamma$-ray data from \citet{yang2025origin}. For these four sources, we use Fermi-LAT data when filtering the simulated fluxes.

\subsection{Filtering}\label{44}

We define the goodness of fit to the neutrino flux by introducing the quantity:
\begin{equation}
\Delta_\nu = \sum_{i} \left[ \max \left(0, \log_{10} \left( \frac{S_{\nu, i}}{U_{\nu, i}} \right) \right) + \max \left( 0, \log_{10} \left( \frac{L_{\nu, i}}{S_{\nu, i}} \right) \right) \right]\,.
\label{eq:photon_filter_result}
\end{equation}

\begin{figure}[h!]
    \centering
    \includegraphics[width=0.8\textwidth]{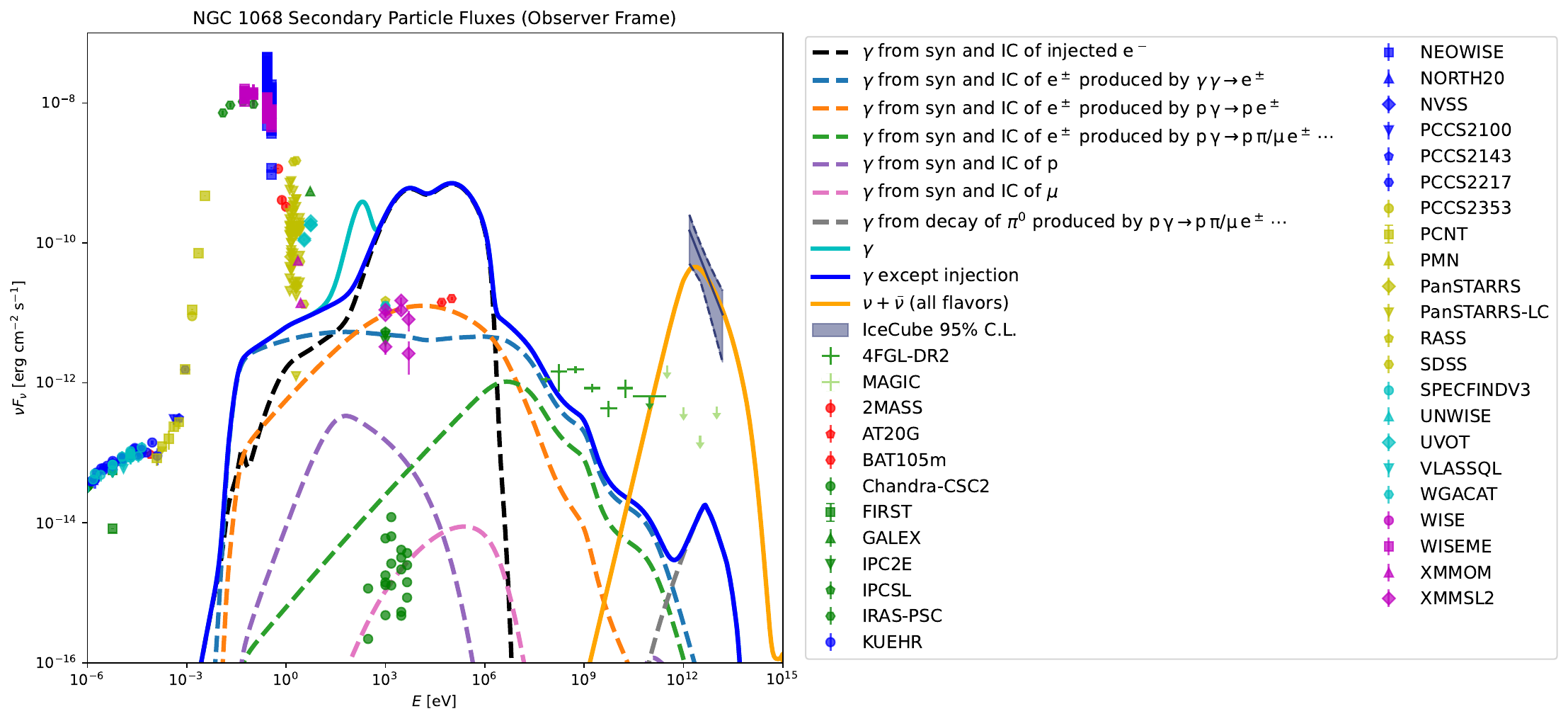}
    \caption{Detailed results for one of the best fits to NGC 1068 data from run A ($p \gamma$ only) with $\Delta \le 0.05$.}
    \label{fig:NGC_1068_run_A_full}
\end{figure}

\begin{figure}[htbp]
    \centering
    \begin{subfigure}[T]{0.3\textwidth}
        \centering
        \includegraphics[width=\textwidth]{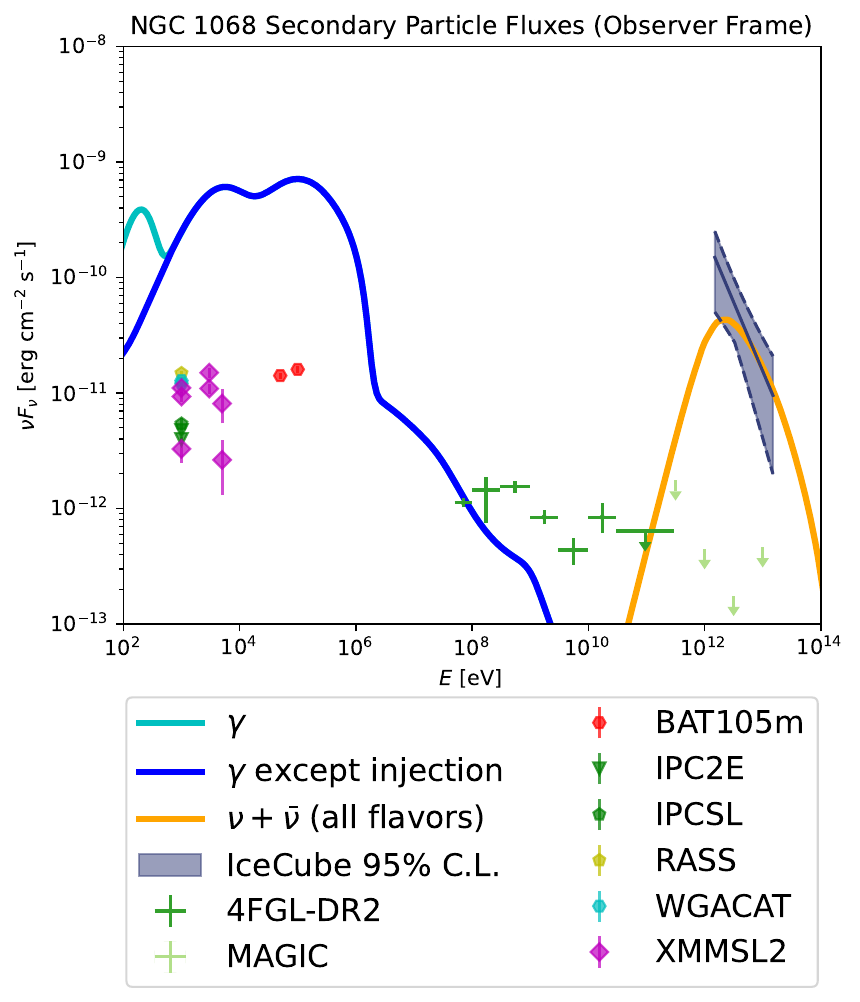}
    \end{subfigure}
    \begin{subfigure}[T]{0.3\textwidth}
        \centering
        \includegraphics[width=\textwidth]{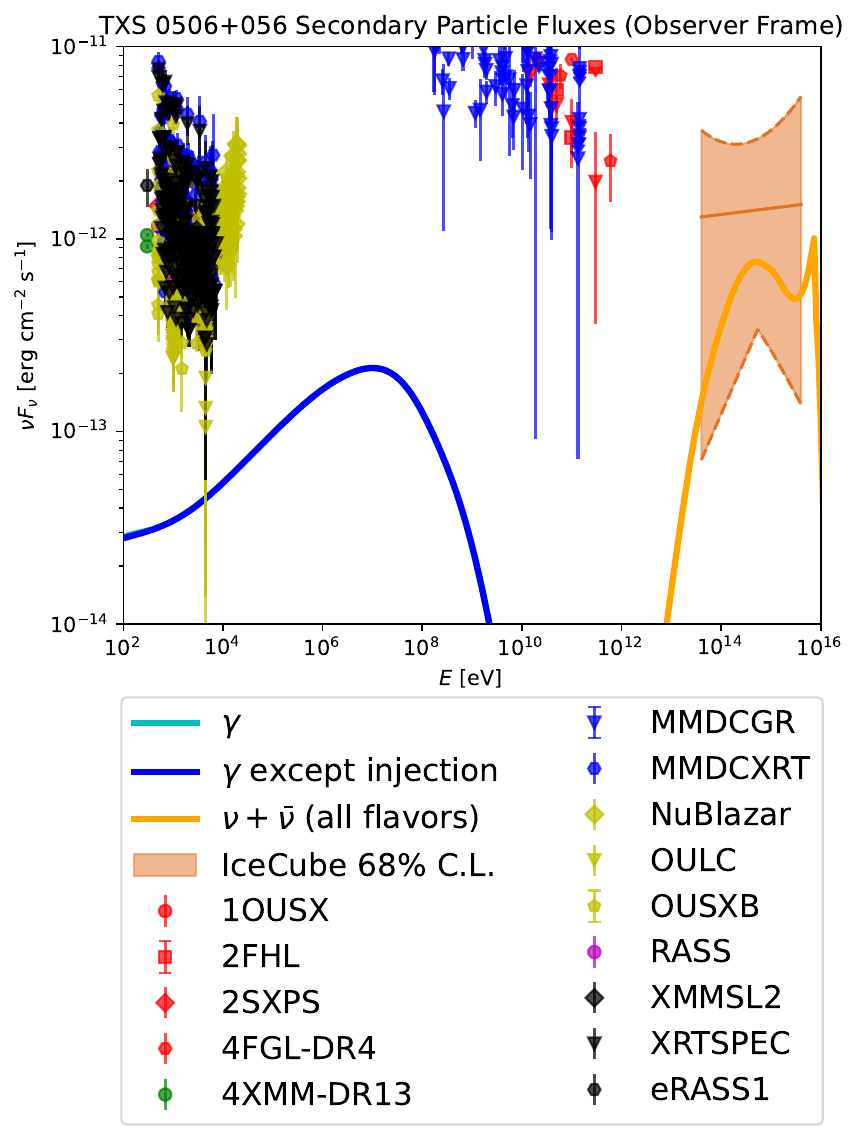}
    \end{subfigure}
    \begin{subfigure}[T]{0.3\textwidth}
        \centering
        \includegraphics[width=\textwidth]{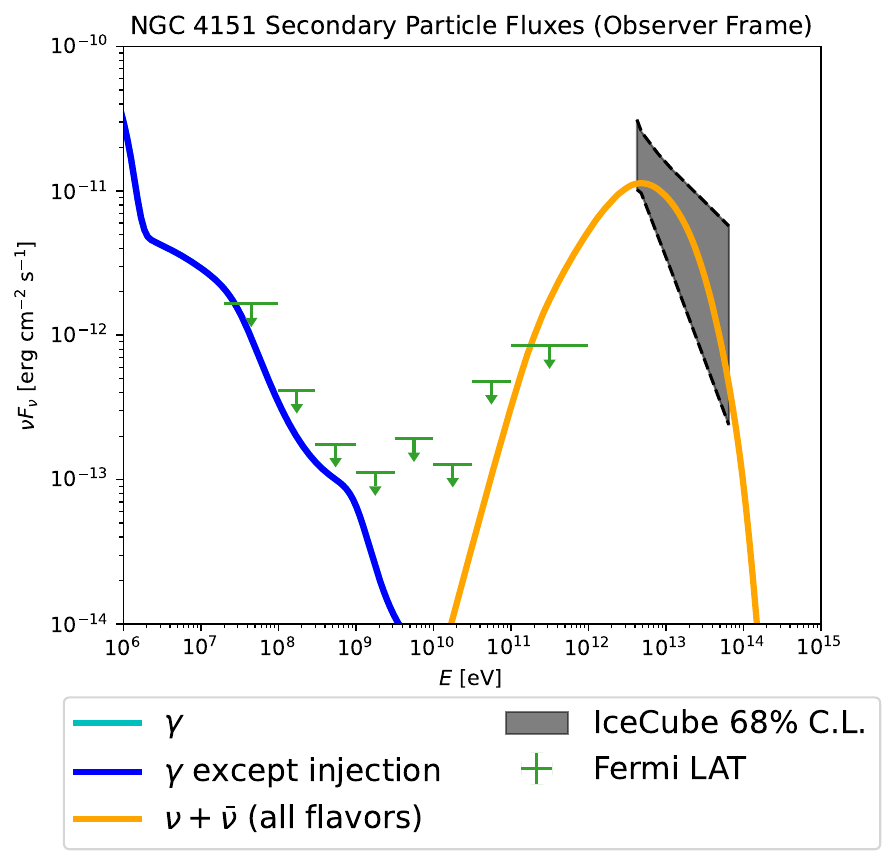}
    \end{subfigure}
    \vspace{1em}
    \begin{subfigure}[T]{0.3\textwidth}
        \centering
        \includegraphics[width=\textwidth]{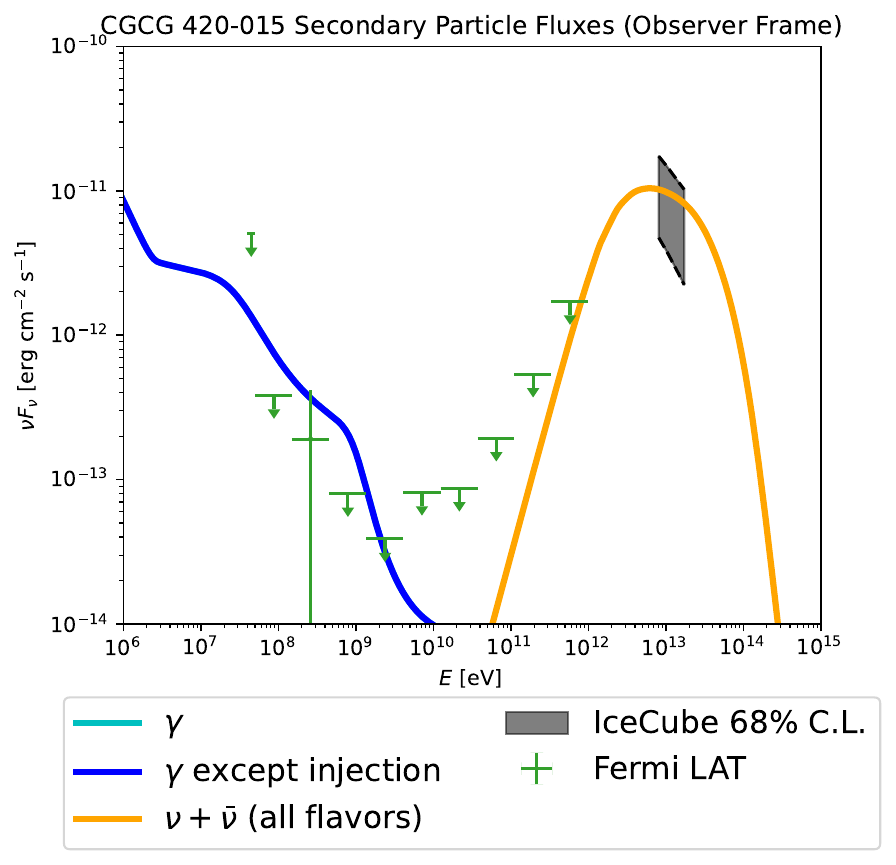}
    \end{subfigure}
    \begin{subfigure}[T]{0.3\textwidth}
        \centering
        \includegraphics[width=\textwidth]{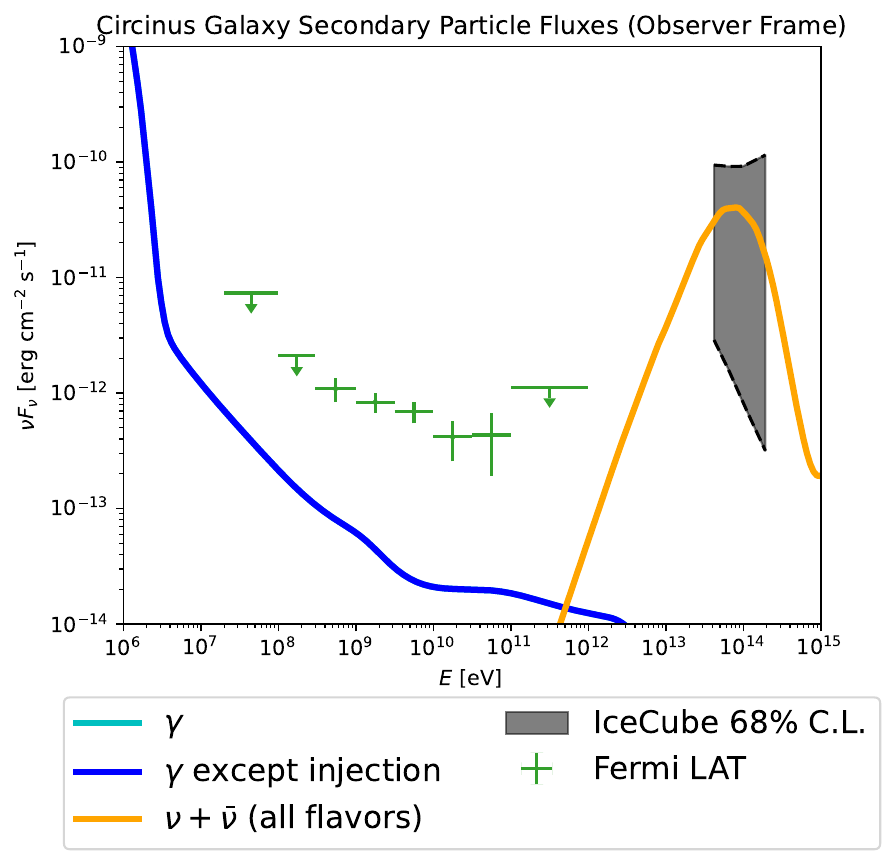}
    \end{subfigure}
    \begin{subfigure}[T]{0.3\textwidth}
        \centering
        \includegraphics[width=\textwidth]{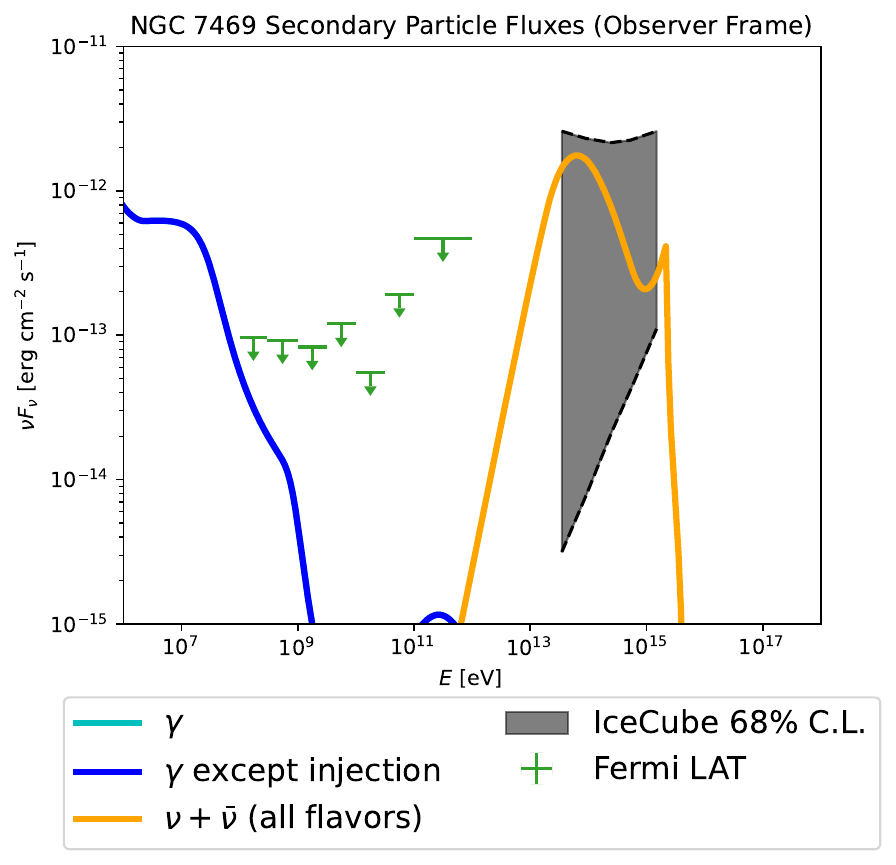}
    \end{subfigure}
    \caption{Simplified results for one of the best fits for each of the six neutrino sources from run A ($p \gamma$ only). All sources are characterized with a $\Delta \le 0.05$, except CGCG 420-015, for which $\Delta = 0.34$.}
    \label{fig:neutrino_sources_run_A_simplified}
\end{figure}

For the photon fluxes, we treat all photon observations as upper limits when filtering the simulations: 
\begin{equation}
\Delta_\gamma = \sum_{i} \max \left( 0, \log_{10} \left( \frac{S_{\gamma, i}}{U_{\gamma, i}} \right) \right),
\end{equation}
where the simulation output is $S_\nu$ for the neutrino flux, $S_\gamma$ for the photon flux. The neutrino flux upper and lower bounds are $U_\nu$ and $L_\nu$, and the photon flux lower bound is $U_\gamma$. The sum is across the energy range of all observations. We then define the average filter result as
\begin{equation}
\Delta = \frac{\Delta_\nu + \Delta_\gamma}{2}.
\end{equation}
Here, $\Delta = 0$ and small values of $\Delta$ indicate good fits to the data. 
For the results shown in Section \ref{sec5}, we list fits with $\Delta \le 0.05$ for a maximum of $1,000$ best fits for each source. We also impose the condition that $R \ge 2 R_S$. We show the detailed results for NGC 1068 in Figure \ref{fig:NGC_1068_run_A_full} and simplified results for other five sources in Figure \ref{fig:neutrino_sources_run_A_simplified}. For CGCG 420-025 in run A, the best fit has $\Delta = 0.34$. The fit is, however, satisfactory as shown in Figure \ref{fig:neutrino_sources_run_A_simplified}. The fit suggests that the Fermi data in the $\rm 100\,MeV \sim 1\,GeV$ range may originate from the photon flux of $\pi^0$ origin.

NGC 1068 is a Compton-thick galaxy with a column density of $\rm N_H = 10^{25} \, \mathrm{cm}^{-2}$\,\citep{marinucci2015nustar} that absorbs X-rays, reducing the observed flux relative to the flux in the core. It has a total cross section $\sigma_{\mathrm{tot}} (E_\gamma)$ and consists of two parts:
\begin{equation}
\sigma_{\mathrm{tot}}(E_\gamma) = \sigma_{\mathrm{KN}}(E_\gamma) + \sigma_{\mathrm{abs}}(E_{\gamma}),
\end{equation}
where $E_\gamma$ is the energy of the X-ray, $\sigma_{\mathrm{KN}}(E_\gamma)$ is the Klein-Nishina cross section\,\citep{yaqoob2012nature} for Compton scattering, and $\sigma_{\mathrm{abs}} (E_\gamma)$ is the absorption cross section. The fraction of X-rays transmitted is given by
\begin{equation}
e^{-\tau (E_\gamma)} = e^{- \sigma_{\mathrm{tot}} (E_\gamma) {\rm N_H}} < e^{- \sigma_{\mathrm{KN}} (E_\gamma) {\rm N_H}} < e^{- \sigma_{\mathrm{KN}} (E_\gamma = 100 \, \mathrm{keV}) {\rm N_H}} \approx e^{- 0.7 \sigma_{\mathrm{T}} {\rm N_H}} \simeq 0.95 \%
\end{equation}
for X-rays with $E_\gamma \lesssim 100 \, \mathrm{keV}$, where $\sigma_{\mathrm{T}} = 6.65 \times 10^{-25} \, \mathrm{cm}^2$ is the Thompson cross section. In this Compton-thick environment, X-rays will be down-scattered multiple times and be finally absorbed by the photo-electric effect. For simplicity, we assume that $99 \%$ of X-rays with energies $100 \, \mathrm{keV}$ and below are absorbed. Therefore, we increase the upper limit $U_\gamma$ in Eq.~\ref{eq:photon_filter_result} to the observed value multiplied by a factor of $100$. We assume that there is no scattering or absorption of $\gamma$-rays. The X-ray flux at the core of NGC 1068 is consistent with our estimates and with the value directly observed by the NuSTAR experiment\,\citep{marinucci2015nustar} during a temporary decrease of the column density of the obscuring material along the line of sight.

For TXS 0506+056, we assume the Galactic value of column density $\rm N_H = 1.11 \times 10^{21} \, \mathrm{cm}^{-2}$~\citep{icecube2018neutrino} and therefore neglect the scattering and absorption of X-rays and $\gamma$-rays. For NGC 4151, CGCG 420-015, Circinus Galaxy, and NGC 7469, we are only fitting $\gamma$-ray observations and similarly ignore scattering and absorption resulting from $\rm N_H$.

\section{Multimessenger Description of IceCube Sources: Results and Analysis}\label{sec5}

One expects that $pp$ interactions contribute at some level to the production of neutrinos. Including them allows for a wider range of fits because of the introduction of an additional parameter $n_p$, but they are not significantly improved. All sources can be described by the interactions of protons on the X-rays in their dense core.

\subsection{Best Fits}\label{sec51}

Figure \ref{fig:NGC_1068_run_A_full} shows the photon and neutrino spectra for one of the best fits to the NGC 1068 data for run A. The X-ray flux shown is the one intrinsic to the the corona before absorption. It is clear that after implementing the absorption as described in Section 2.4, the flux will be consistent with the data. The bulk of the corona X-rays are produced by synchrotron radiation of the injected electron plasma and by their IC scattering on Planck radiation from the accretion disk. The X-rays efficiently absorb the $\gamma$-rays produced in association with the neutrinos by pair production. The electron-positron pair produced in this process yields more X-rays by the synchrotron and IC processes, as shown in the figure.

For comparison, Figure~\ref{fig:neutrino_sources_run_A_simplified} shows one of the best fits to six sources in run B, which allow for $pp$ interactions. These represent a subdominant process for all sources. For TXS 0506+056, X-rays and $\gamma$-ray observations are not subjected to strong constraints. TXS 0506+056 is a blazar; X-rays and $\gamma$-rays can originate from other sites than the dense corona, most prominently the jet.

\begin{figure}[ht!]
    \centering
    \includegraphics[width=1.0\textwidth]{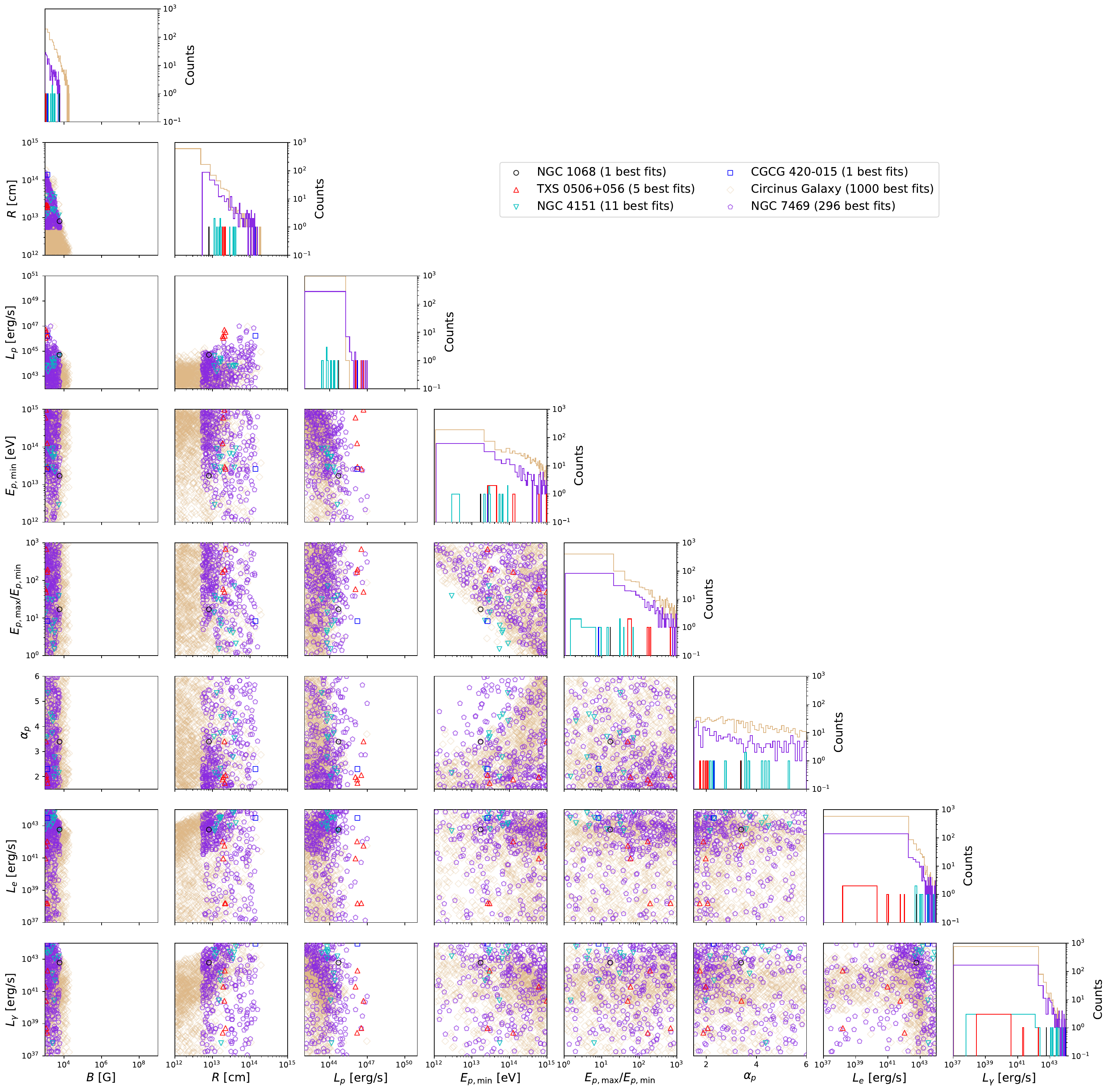}
    \caption{Best fit parameters for the six sources from run A ($p \gamma$ only). The fits are filtered as before. A maximum number of $1,000$ best fits are shown for each source.}
    \label{fig:parameter_correlation_run_A}
\end{figure}

\begin{figure}[ht!]
    \centering
    \includegraphics[width=0.95\textwidth]{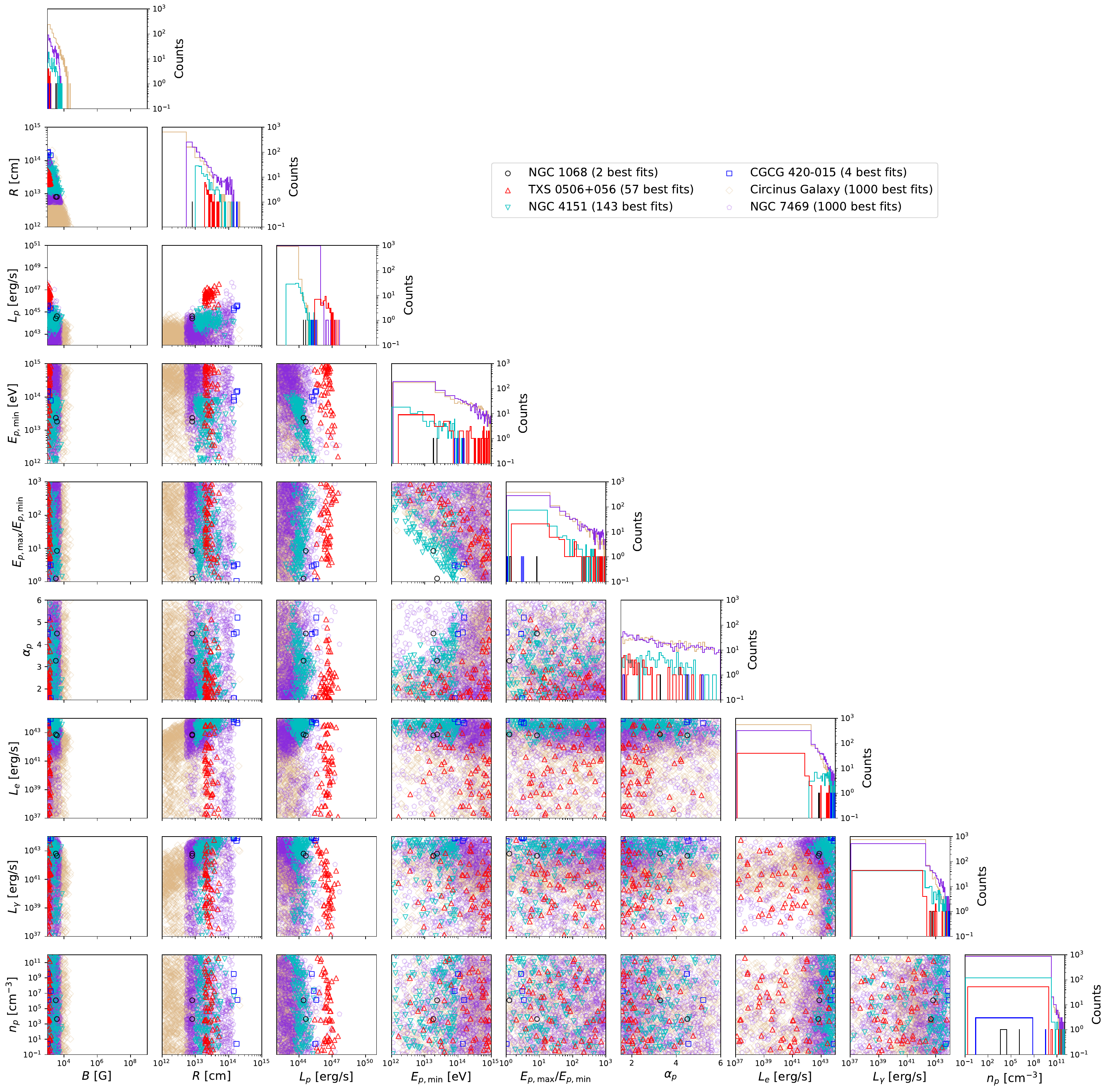}
    \caption{Best fit parameters ($\Delta \le 0.05$) for the six sources from run B ($p \gamma$ and $pp$). A maximum number of $1,000$ best fits are shown for each source.}
    \label{fig:parameter_correlation_run_B}
\end{figure}

The distribution of the parameter values corresponding to the fits are shown in Figures~\ref{fig:parameter_correlation_run_A} and \ref{fig:parameter_correlation_run_B} for run A and run B,  respectively. The observations put an upper limit on the magnetic field $B$, with TXS 0506+056 and CGCG 420-015 both fitting low $B$ values. All sources prefer low $R$ values, in particular NGC 1068 and TXS 0506+056. The figure for run B underscores that NGC 1068 data do not fit to high values of $n_p$.


For each source, the parameters describing the best fits tend to cluster, and there are apparent correlations between the parameters:
\begin{itemize}
    \item For high values of $R$, $B$ is reduced to avoid overproduction of X-rays.
    \item Also for higher $L_p$ values, $B$ must be reduced because otherwise $\gamma$-rays will be overproduced in $p \gamma$ interactions.
    \item For larger $L_p$ values, $R$ has to be reduced. Because the data fits a value for $\frac{dN_p}{dE_p dtdV}$ to the observed neutrino flux, increasing $L_p$ can be compensated by a reduction in volume $V$ and, therefore, $R$.
    \item $E_{p, \mathrm{max}}, E_{p, \mathrm{min}}$ and $\alpha_p$ are all correlated for the same reason.
\end{itemize}
%
In general, the fitted parameters exhibit the characteristics of the dimensional analysis of the corona model described in Section\,\ref{sec2}. With the neutrino flux given by $L_\nu = \tau_{p \gamma}\,L_p$, it follows from Eq.\,\ref{eq:tau} that the neutrino flux is proportional to the X-ray flux and inversely proportional to the radius, the two quantities that determine the density of the target. Our fits prefer low values of $\rm R$. As $\rm R$ increases, $L_X$ will increase, pushing the X-ray flux and therefore the average filter result $\Delta$ to higher values, worsening the agreement with the observations.

\begin{figure}[h!]
    \centering
    \begin{subfigure}[T]{0.49\textwidth}
        \centering
        \includegraphics[width=\textwidth]{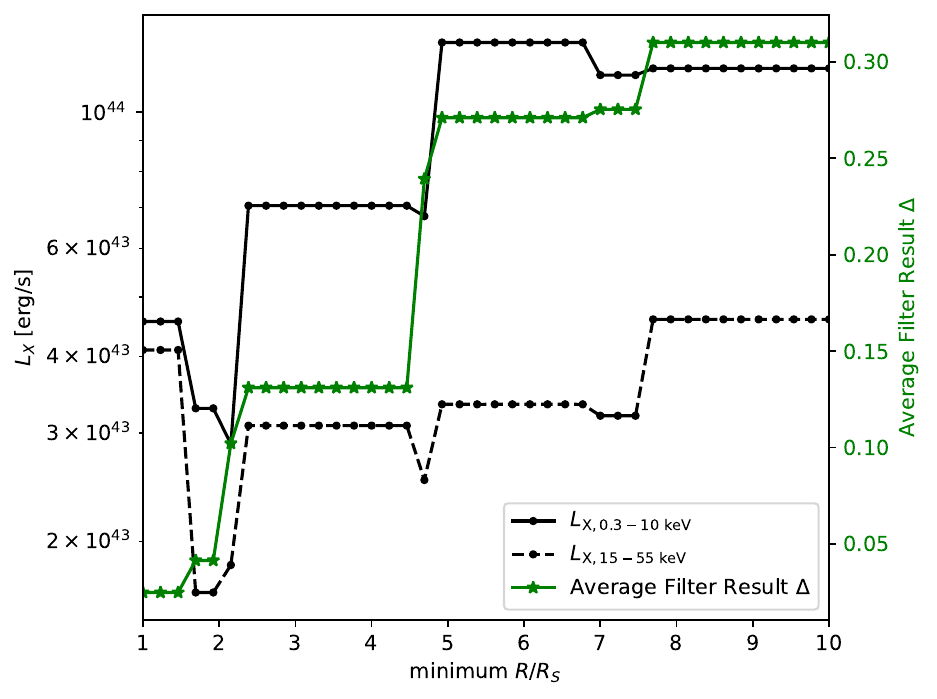}
        \caption{Run A ($p \gamma$ only).}
    \end{subfigure}
    \begin{subfigure}[T]{0.49\textwidth}
        \centering
        \includegraphics[width=\textwidth]{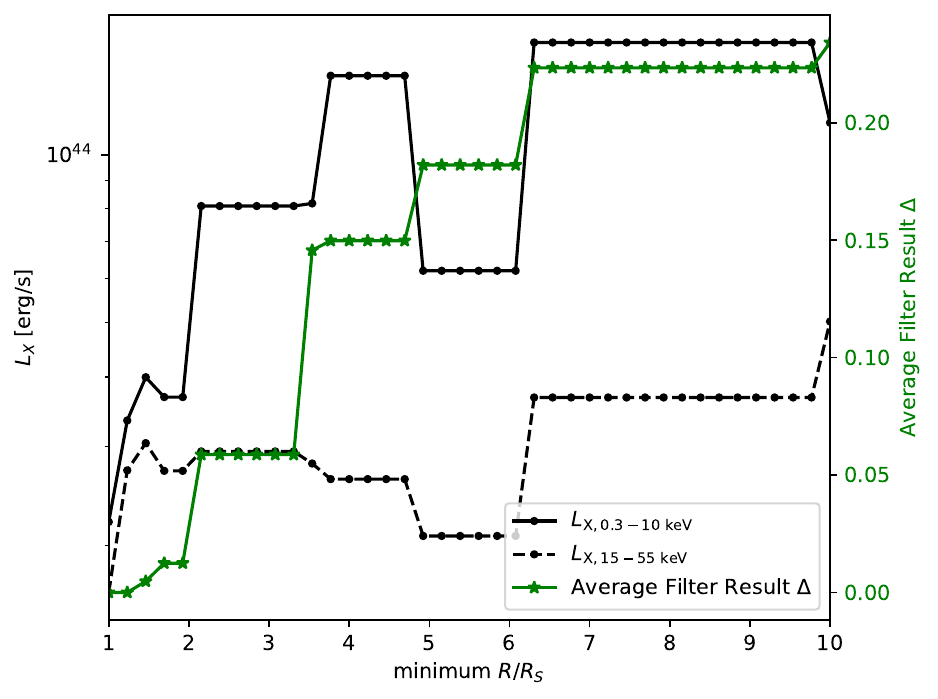}
        \caption{Run B ($p \gamma$ and $pp$).}
    \end{subfigure}
    \caption{\emph{Min} $R - L_X$ dependence for the best fits to the NGC 1068 multimessenger data for (a) run A and (b) run B.}
    \label{fig:minR_LX}
\end{figure}

The observations of NGC 1068 show a preference for lower $\rm R$ values in simulations. To further investigate this, we gradually increase the minimum $\rm R$ value and refit the data. The resulting values of the X-ray luminosity $L_X$ are plotted in Figure~\ref{fig:minR_LX} for run A and run B, displaying the preference for the lower values of the corona radius in terms of the goodness of fit $\Delta$. The tendency of $\rm R$ towards low values is of course enhanced by the compact geometry that we assumed; for more realistic geometries of the corona, like a lamppost geometry, the realistic volume is likely increased. 

In summary, we test the linear relation between the values of $L_\nu$, $L_X$ and $L_p$ in our model. We study this relation in terms of flux rather than luminosity in order to remove the bias that sources at larger distances have reduced fluxes, a correlation that biases the dynamical correlation we want to investigate:
\begin{equation}
F = \frac{L}{4 \pi d_L^2},
\end{equation}
where $d_L$ is the luminosity distance from each source.

\begin{figure}[htbp]
    \centering
    \begin{subfigure}[T]{0.35\textwidth}
        \centering
        \includegraphics[width=\textwidth]{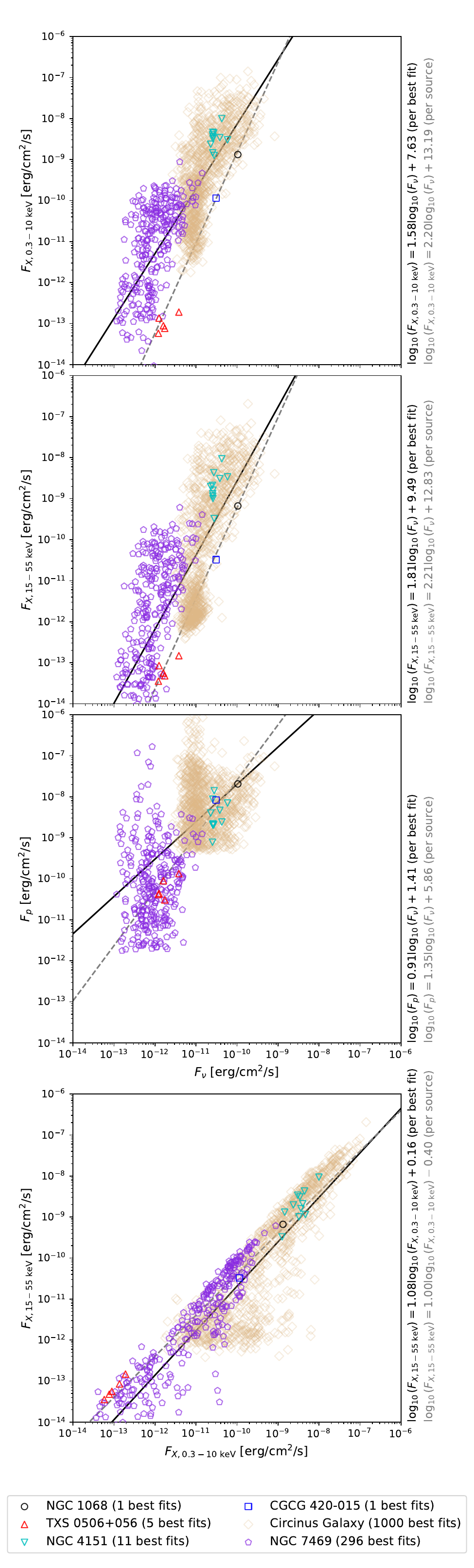}
        \caption{Run A ($p \gamma$ only).}
    \end{subfigure}
    \begin{subfigure}[T]{0.35\textwidth}
        \centering
        \includegraphics[width=\textwidth]{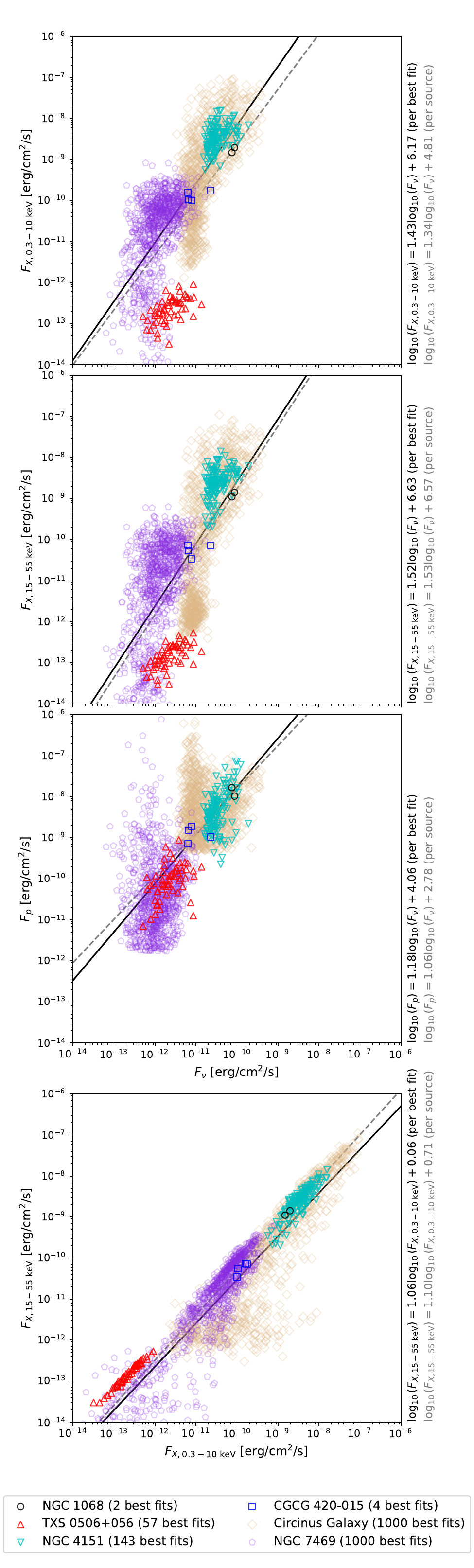}
        \caption{Run B ($p \gamma$ and $pp$).}
    \end{subfigure}
    \caption{$F_{\nu}$ and $F_p$ dependence on $F_X$ for the best fits to the sources for (a) run A and (b) run B. The fits are filtered as before. A maximum number of $1,000$ best fits are shown for each source. The bottom two plots show the correlation between the X-ray fluxes in low and high X-ray energy ranges for our modeling of the corona.}
    \label{fig:Fnu_FX_Fp}
\end{figure}

As shown in Figure \ref{fig:Fnu_FX_Fp}, runs A and B exhibit similar clustering for each source and similar slope. We fit these data using two methods: per best fit (an equally weighted average of all best fits across all sources) and per source (an average of the best fits within each source, followed by an average across sources).

There is a linear correlation between $F_X$ and $F_\nu$, with a slope that favors the $F_X$ axis. For run A we have:
\begin{align}
    \log_{10}(F_{X, \mathrm{0.3 - 10\ keV}}) &= 1.58 \log_{10}(F_\nu) + 7.63 \, \, \, \, (\mathrm{per \, best \, fit}), \\
    \log_{10}(F_{X, \mathrm{0.3 - 10\ keV}}) &= 2.25 \log_{10}(F_\nu) + 13.79 \, (\mathrm{per \, source}), \\
    \log_{10}(F_{X, \mathrm{15 - 55\ keV}}) &= 1.80 \log_{10}(F_\nu) + 9.37 \, \, \, \, (\mathrm{per \, best \, fit}), \\
    \log_{10}(F_{X, \mathrm{15 - 55\ keV}}) &= 2.28 \log_{10}(F_\nu) + 13.64 \, (\mathrm{per \, source}).
\end{align}
And similarly for run B:
\begin{align}
    \log_{10}(F_{X, \mathrm{0.3 - 10\ keV}}) &= 1.42 \log_{10}(F_\nu) + 5.92 \, (\mathrm{per \, best \, fit}), \\
    \log_{10}(F_{X, \mathrm{0.3 - 10\ keV}}) &= 1.32 \log_{10}(F_\nu) + 4.60 \, (\mathrm{per \, source}), \\
    \log_{10}(F_{X, \mathrm{15 - 55\ keV}}) &= 1.50 \log_{10}(F_\nu) + 6.32 \, (\mathrm{per \, best \, fit}), \\
    \log_{10}(F_{X, \mathrm{15 - 55\ keV}}) &= 1.52 \log_{10}(F_\nu) + 6.41 \, (\mathrm{per \, source}).
\end{align}
Also, $F_p$ remains above $F_\nu$, which implies that active galaxies do not behave as calorimeters, as protons deposit only a small fraction of their energy in the coronal region to produce neutrinos. Finally, the reason $F_{X, \mathrm{0.3 - 10\ keV}} \sim F_{X, \mathrm{15 - 55\ keV}}$ leads to a similar conclusion follows from the fact that all sources prefer similar values of the magnetic field $B$, and therefore have a similar ratio in the power of the synchrotron radiation and IC processes. The linear dependence between $F_p$, $F_X$, and $F_\nu$ apparent in our corona model description of the sources has been anticipated in the papers of Kun and collaborators\,\citep{kun2024correlation}. They argued for a correlation with higher energy values of the X-ray flux, but this is not critical.

\subsection{Final Considerations}\label{sec52}

\begin{figure}[ht!]
    \centering
    \begin{subfigure}[T]{0.49\textwidth}
        \centering
        \includegraphics[width=\textwidth]{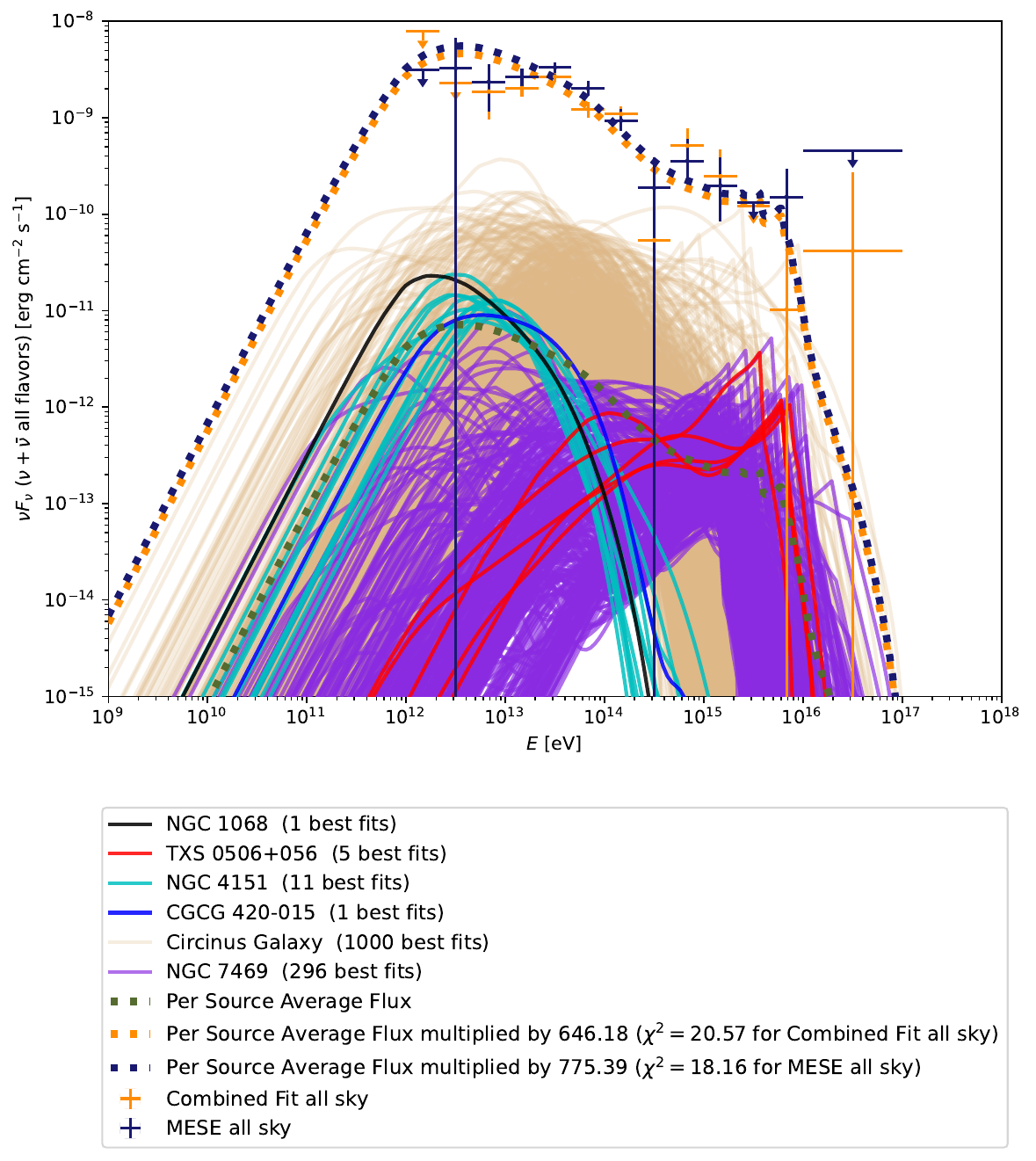}
        \caption{Run A ($p \gamma$ only).}
    \end{subfigure}
    \begin{subfigure}[T]{0.49\textwidth}
        \centering
        \includegraphics[width=\textwidth]{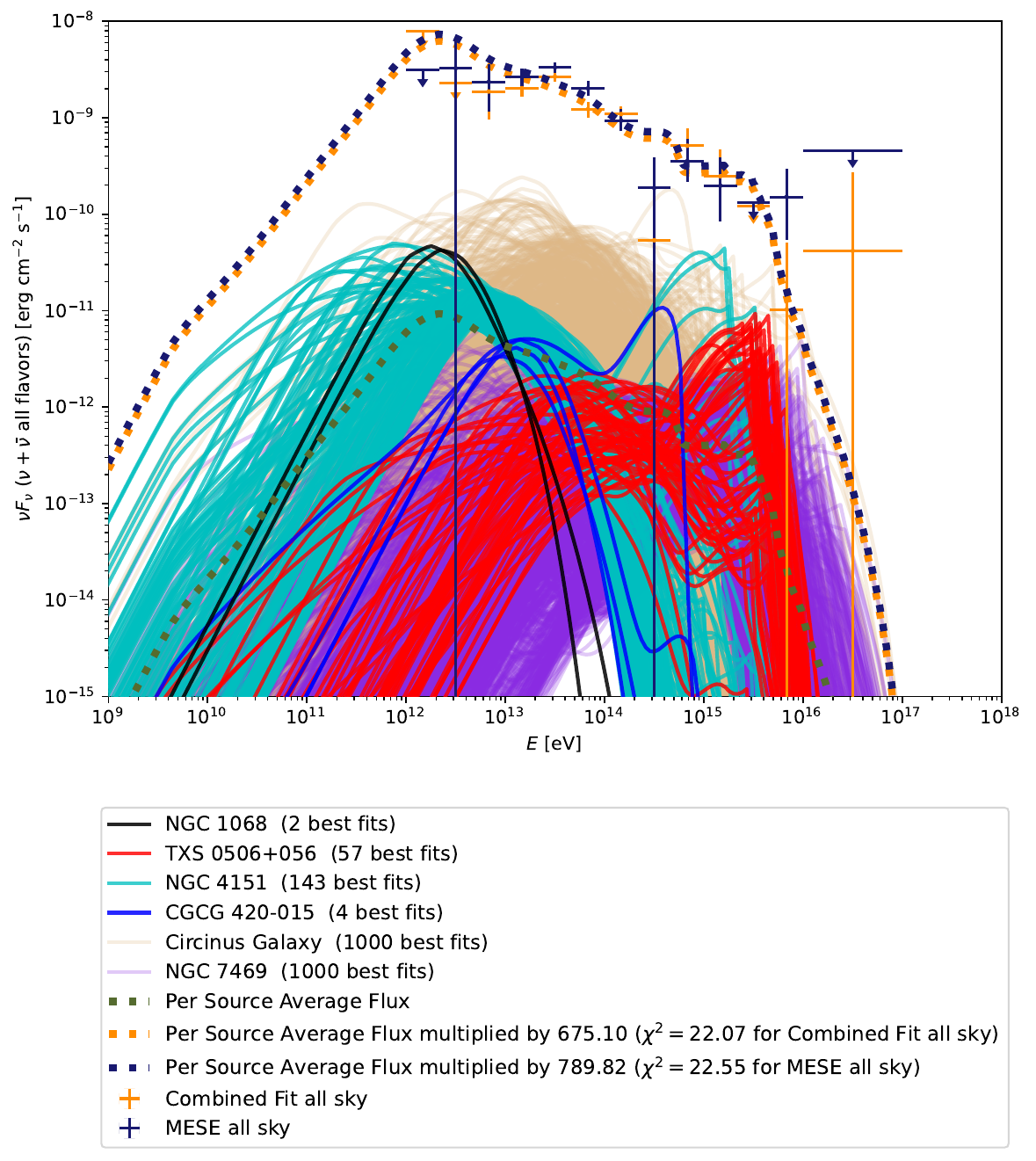}
        \caption{Run B ($p \gamma$ and $pp$).}
    \end{subfigure}

    \caption{Fitting the extragalactic diffuse neutrino flux with the best fits. The fits are filtered as before. A maximum number of $1,000$ best fits are shown for each source.}
    \label{fig:diffuse_flux_fitting}
\end{figure}

We conclude that the top sources in the IceCube neutrino sky are active galaxies with a dense core emitting X-rays. We do not think that it matters much whether they are jetted or not. They may have jets, and these may even emit gamma rays causally connected to the time that neutrinos were produced, as discussed in the introduction, but they are not the site where the neutrinos are produced, close to the central black hole. Where the energy in the associated gamma rays emerges after absorption in the core and in the EBL is not clear and depends sensitively on the determination of the diffuse flux below the feature at $\sim 30 \, \mathrm{TeV}$; for a recent discussion, see Ref.~\citet{KhateeZathul:2026oam}.

We close by estimating the number of neutrino-emitting active galaxies, $N_{\mathrm{AGN}}$, required to describe the diffuse extragalactic flux, assuming that the set of sources we considered is representative of sources producing cosmic rays and assuming a uniform distribution in luminosity distance. We average the flux of the sources in the observer frame and rescale it by a factor $N_{\mathrm{AGN}}$, which accommodates the diffuse flux. When averaging the flux from sources, we average the best fits in one source first and then average over sources. Figure \ref{fig:diffuse_flux_fitting} shows the results for runs A and B. The IceCube ``combined fit" and ``medium energy starting events" (MESE) data sets refer to the two different measurements of the diffuse cosmic neutrinos flux described in references \citet{abbasi2026evidence} and \citet{abbasi2026improved}. We multiplied source fluxes by $4 \pi$ to obtain their contribution to the all sky flux.

Both runs A and B require $N_{\mathrm{AGN}} \sim 650$ to $800$ to fit the diffuse flux. Adding $pp$ interactions does not significantly change the result. The reasons are that the spectral break at $\sim 30 \, \mathrm{TeV}$ is a direct result of $\Delta$ resonance and that $pp$ interaction is not the main contribution of the neutrino flux\,\citep{KhateeZathul:2026oam}. The fact that our number of sources agrees with the number of active galaxies with an X-ray flux similar to or larger than the one of NGC 1068\,\citep{Anchordoqui:2021vms}, about $800\, \mathrm{Gpc}^{-3}$, may not be a coincidence. It is also possible that at the highest energies cosmic-ray acceleration may require larger structures à la Hillas, which would naturally be clusters and superclusters within our proposal. For these, the neutrino-producing target material may be less dense, and it will take next-generation detectors to probe the neutrinos they produce.

\begin{acknowledgments}
This work used the following software: Astrophysical Multimessenger Modeling (AM3)\,\citep{klinger2024am3}, Astropy\,\citep{astropy:2013, astropy:2018,
  astropy:2022}, and Gammapy v1.2\,\citep{gammapy:2023, gammapy:zenodo-1.2}. It has been supported by the Office of the Vice Chancellor for Research at the University of Wisconsin--Madison with funding from the Wisconsin Alumni Research Foundation.  F.H. and C.L. are also supported by the National Science Foundation under grants~PHY-2209445 and OPP-2042807. The authors thank Anatoli Fedynitch, Carlos Argüelles-Delgado, and Ali Kheirandish for their helpful discussions and constructive comments on this work.
\end{acknowledgments}


\bibliography{Li_Halzen_article}{}
\bibliographystyle{aasjournalv7}



\end{document}